\RequirePackage{ifpdf}

\documentclass[preprint,nofootinbib,tightenlines,nobibnotes]{revtex4-1}

\ifpdf
  \usepackage{color}
\fi

\usepackage[latin1]{inputenc}
\usepackage{amsmath}
\usepackage{amsfonts}
\usepackage{amssymb}
\usepackage{float}
\usepackage{epsfig}
\usepackage{mathrsfs}

\ifpdf
  \usepackage{graphicx}
  \usepackage[unicode]{hyperref}
\else
  \usepackage{graphics}
\fi

\newcommand{\td}{\mathrm{d}}

\DeclareMathOperator{\AdS}{AdS}

\DeclareMathOperator{\vol}{vol}
\newcommand{\const}{\text{const}}

\newcommand{\Lie}[1]{{#1}}

\newcommand{\re}{\text{Re\;}}
\newcommand{\im}{\text{Im\;}}

\begin{document}

\preprint{IPMU12-0006}

\title{On supersymmetry breaking three-form flux on Sasaki-Einstein
  manifolds}

\author{Johannes Schmude}
\email{johannes.schmude@ipmu.jp}
\affiliation{Institute for the Physics and Mathematics of the Universe\\
The University of Tokyo, Kashiwa, Chiba 277-8586, Japan}

\date{\today}

\begin{abstract}
  Studying nonsupersymmetric yet imaginary self-dual three-form fluxes
  in type IIB supergravity backgrounds on Sasaki-Einstein manifolds we
  find a new analytic solution that restores supersymmetry in the IR,
  breaks it at higher energies, yet suffers from curvature
  singularities in the UV, when a certain SUSY-breaking parameter
  becomes large.

  Consequently we investigate a variety of possibilities to cure the
  singularity by either introducing additional sources, changing the
  fluxes or deforming the geometry. Since it is not possible to cure
  the singularities making physically reasonably assumptions, we
  suggest that there might be a no-go theorem disallowing such
  flux.
\end{abstract}

\maketitle

\section{Introduction}
\label{sec:introduction}

Progress in gauge/string duality is considerably helped by
supersymmetry in various guises: The construction and
deformation of supergravity backgrounds is often based on the study of
the relevant BPS-equations. D-branes and their backreaction can be
addressed using $\kappa$-symmetry or generalized
calibrations. Four-dimensional geometries preserving four supercharges
can be described in terms of generalized geometry. Finally,
the stability of both the backgrounds and embedded branes follows
directly from the existence of conserved supercharges. Yet albeit the
uses of supersymmetry, there is an obvious interest in gauge/string
duality for non-supersymmetric solutions.

Of course, the problem is not a new one and there have been a number
of different approaches. Examples of explicit supersymmetry-breaking
can be found in \cite{Witten:1998zw} and the Sakai-Sugimoto model
\cite{Sakai:2004cn}, which are both based on D4-branes wrapping an
$S^1$. Supersymmetry is broken due to antiperiodic boundary conditions for
the fermions on the $S^1$. Conversely, models such as
\cite{Khavaev:1998fb} and \cite{Gubser:1999pk} are based on
deformations of $\AdS_5 \times S^5$. By giving a radial profile to
various scalar functions in either five- or ten-dimensional
supergravity they allow in principle for both supersymmetric and
non-supersymmetric solutions. Another approach is to study linearized
perturbations around a known supersymmetric solutions, as was done in
\cite{Aharony:2002vp} for the Maldacena-N\'u\~nez background and in
\cite{Borokhov:2002fm} and \cite{Kuperstein:2003yt} for the
Klebanov-Strassler solution. Further progress has been made in the
context of the latter. Here, \cite{Dymarsky:2009cm} is concerned with
a modification of the calibration condition for D7-branes, while there
has been very recent progress towards incorporating the backreaction of
anti-branes, see e.g.~\cite{Bena:2011wh}.

The purpose of this note is to investigate the possibility of
following a different approach. The various conifold theories (see
\cite{Klebanov:1998hh}, \cite{Klebanov:2000nc} 
and \cite{Klebanov:2000hb}), their generalizations to metric cones
over Sasaki-Einstein manifolds (\cite{Gauntlett:2004yd},
\cite{Martelli:2004wu}) as well as the Maldacena-N\'u\~nez background
\cite{Maldacena:2000yy} are all examples of $\Lie{SU}(3)$-structure
backgrounds in type IIB string theory. Hence one can ask whether it is
possible to exploit the existence of the $G$-structure when breaking
the supersymmetry in these solutions.

In the context of string compactifications this question has been
addressed by the Domain Wall Supersymmetry Breaking (DWSB) mechanism
of \cite{Lust:2008zd}. The idea is as follows. Warped supergravity
backgrounds of the form $\mathbb{R}^{1,3} \times_A \mathcal{M}_6$
preserving four supercharges can be described in terms
of generalized geometry and an $\Lie{SU}(3) \times
\Lie{SU}(3)$-structure on the generalized tangent bundle -- see
\cite{Grana:2004bg}. Thus, a supersymmetric solution of the equations
of motion can then be constructed following three steps. Once one has
established the existence of the $G$-structure, one imposes certain
differential conditions taking the role of BPS-equations on it and
finally studies the Bianchi identities for possible sources. Now, the
authors of \cite{Lust:2008zd} observed that it is possible to 
change the differential conditions while maintaining the existence of
the $G$-structure.\footnote{
When thinking of supersymmetry in terms of spinors, this means that
there is still a globally non-vanishing spinor $\epsilon$ that does
however no longer satisfy the supersymmetry equations $\delta_\epsilon
\lambda = 0$ and $\delta_\epsilon \psi_\mu = 0$ for dilatino and
gravitino.
}
Subsequently, one obtains non-supersymmetric solutions
to the full equations of motion. Supersymmetry breaking is
characterized by a complex function $\mathfrak{r} : \mathcal{M}_6 \to
\mathbb{C}$. The special case $\mathfrak{r} = 0$ restores the
differential conditions (and thus the solution) to the original SUSY
ones satisfied by the geometries we mentioned previously.

As we will briefly discuss in section \ref{sec:supergravity}, solving
the DWSB solutions in full generality is still a formidable 
problem. Hence we will subsequently restrict our attention to
Sasaki-Einstein geometries, where it is possible to restrict them
further to the case of imaginary self-dual three-form flux that was
already studied in \cite{Giddings:2001yu} and the starting point of
the DWSB paper. Note that \cite{Kuperstein:2003yt} discusses
nonsupersymmetric imaginary self-dual flux on the deformed conifold.

The outline of this paper is as follows: After introducing the DWSB
equations in section \ref{sec:supergravity} and presenting the
relevant Sasaki-Einstein geometries in \ref{sec:examples}, we will
spend the bulk of the paper discussing the possibility of identifying
the NS three-form with the $(3,0)$-form $\Omega$ that characterizes the
$\Lie{SU}(3)$-structure. We find an analytic solution which
breaks SUSY in the UV and restores it in the IR, yet is
plagued by a naked singularity in the UV. We subsequently investigate
the possibility of curing the singularity by further modifying the
solution (section \ref{sec:no_way_out}). Here, we consider the
addition of D-brane and orientifold sources following
\cite{Benini:2006hh} and that of additional, supersymmetric flux. The
latter allows us to contrast our ansatz with the Klebanov-Tseytlin
\cite{Klebanov:2000nc} solution and its generalization
\cite{Herzog:2004tr}. In the appendices we summarize some conventions
as well as some technical details concerning the energy-momentum
tensor.

\section{The DWSB equations in type IIB}
\label{sec:supergravity}

Let us start by summarizing a few essentials of type IIB supergravity
in the democratic formalism. Except where explicitly noted, we will
work in string frame. We discuss the essential ingredients here and
refer more technical aspects to appendix \ref{sec:conventions}. In the
democratic  formalism of \cite{Bergshoeff:2001pv}, the $RR$ field
strengths are doubled to $\{ F_1, F_3, F_5, F_7, F_9 \}$. The action is
\begin{equation}
  \begin{aligned}
    S_s &= \frac{1}{2\kappa_{10}^2} \int \td^{10}x \sqrt{-g} \left\{
      e^{-2\Phi} \left\lbrack R + 4 \partial\Phi^2 - \frac{1}{2} H^2
      \right\rbrack - \frac{1}{4} F^2 \right\}
  \end{aligned}
\end{equation}
The sum over the different RR forms is implicit. Due to the doubling
of $RR$ fields, the Bianchi identities and RR equations of motion can
be conveniently summarized in terms of polyforms as
\begin{equation}\label{eq:Bianchi_identity}
   \td_H F = -j
\end{equation}
Here $j$ parametrizes possible sources. Details on the action
used for the sources are relegated to appendix \ref{sec:type-iib}. The
twisted exterior derivative $\td_H$ is defined in equation
\eqref{eq:twisted_exterior_derivative}.

We are interested ten-dimensional spaces that are a warped product of
the form $\mathbb{R}^{1,3} \times_A \mathcal{M}_6$. That is, the
metric is of the form
\begin{equation}\label{eq:DWSB_metric}
  \td s_{10}^2 = e^{2A} \td x^\mu \td x_\mu + g_{ij} \td y^i \td y^j 
\end{equation}
while Poincar\'e invariance dictates the fluxes to take the
form\footnote{It follows that
  \begin{equation*}
    \begin{aligned}
      F_1 &= - \star_6 \tilde{F}_5 &
      F_3 &= \star_6 \tilde{F}_3 &
      F_5 &= - \star_6 \tilde{F}_1
    \end{aligned}
  \end{equation*}
}
\begin{equation}
  \begin{aligned}
    F = F + \td x^0 \wedge \td x^1 \wedge \td x^2 \wedge \td x^3
    \wedge e^{4A} \tilde{F} \qquad
    \tilde{F} = \tilde{\star}_6 F
  \end{aligned}
\end{equation}
Finally we assume the existence of an $\Lie{SU}(3)$-structure which is
defined in terms of an almost complex structure $J$ and an associated
$(3,0)$-form $\Omega$ that satisfy
\eqref{eq:SU_3-structure_normalization}.

With all conventions set up, it is time to turn to the pseudo-BPS
equations of the DWSB formalism \cite{Lust:2008zd}. First of all there
are two equations that are identical to the supersymmetric case:
\begin{subequations}\label{eq:bps-ness_equations}
  \begin{align}
    \td_H (e^{4A-\Phi} \re e^{\imath \theta} e^{\imath J}) &= e^{4A}
    \tilde{F} \label{eq:bps-ness_Flux} \\
    \td_H (e^{2A-\Phi} \im e^{\imath \theta} e^{\imath J}) &= 0 \label{eq:bps-ness_BPS}
 \end{align}
\end{subequations}
For supersymmetric backgrounds the set of BPS equations is completed
by 
\begin{equation}
  \td_H (e^{3A-\Phi} e^{-\imath\theta} \Omega) = 0
\end{equation}
However, in the
DWSB case it is this equation that gets modified. In full generality
\begin{equation}\label{eq:non-bps-ness_equation}
  \begin{aligned}
    \td_H (e^{3A-\Phi} e^{-\imath\theta} \Omega) &= - 4 \imath
    \mathfrak{r} e^{3A-\Phi} \frac{\sqrt{\det g\vert_\Pi}}{\sqrt{\det
        g\vert_\Pi + R}} e^{-R} \wedge \sigma (\vol_\perp)
  \end{aligned}
\end{equation}
Here, $\Pi$ denotes a split of the tangent bundle $T_*\mathcal{M}_6 =
T_*\Pi \oplus T_*^\perp\Pi$ that also leads to a decomposition of the
volume form $\vol_6 = \vol_\Pi \wedge \vol_\perp$, while $R$ is a real
two-form on $\Pi$: $R \in \bigwedge^2 T^*\Pi$. Most importantly, the
complex function $\mathfrak{r}$ parametrizes the supersymmetry
breaking -- obviously, for $\mathfrak{r} = 0$ one recovers the
supersymmetric case. There are two further conditions given in
\cite{Lust:2008zd}. However, since the equations are rather
complicated yet in our case automatically satisfied, we relegate them
to appendix \ref{sec:full-set-dwsb} -- see equations
\eqref{eq:b-field_eq} and \eqref{eq:einstein_eq}.

The Sasaki-Einstein geometries we are concerned with all satisfy
$\theta = 0$ in \eqref{eq:bps-ness_equations} and
\eqref{eq:non-bps-ness_equation}, while $\theta = \frac{\pi}{2}$
holds for Maldacena-N\'u\~nez-like geometries. Hence we set $\theta$
to zero and thus obtain the following equations from
\eqref{eq:bps-ness_BPS}:
\begin{equation}
  \begin{aligned}
    0 &= \td(e^{2A-\Phi}J) &
   0 &= H \wedge J
  \end{aligned}
\end{equation}
The RR-fluxes are fully determined by \eqref{eq:bps-ness_Flux}:
\begin{equation}
  \begin{aligned}
    F_5 &= - e^{-4A} \star_6 \td(e^{4A-\Phi}) \\
    F_3 &= e^{-\Phi} \star_6 H \\
    F_1 &= \frac{e^{-4A}}{2} \star_6 \td(e^{4A-\Phi}J \wedge J)
  \end{aligned}
\end{equation}

\section{Sasaki-Einstein manifolds}
\label{sec:examples}

Before continuing to study these equations, we briefly introduce two
examples of Sasaki-Einstein geometry that we will use later on for
explicit calculations. Our discussion follows \cite{Martelli:2004wu} and
\cite{Gauntlett:2004yd}. Note that we rescale the internal part of
the metric \eqref{eq:DWSB_metric},
\begin{equation}
  \td s_{10}^2 = e^{2A} \td x^\mu \td x_\mu + e^{-2A} \td s_6^2 
\end{equation}
In general we are interested in the case where $\mathcal{M}_6$ is a
cone over a Sasaki-Einstein manifold. In other words, the metric can
be written locally as
\begin{equation}
  \td s_6^2 = \td r^2 + r^2 \left\lbrack \td s_4^2 + \left(
      \frac{\td\psi^\prime}{3} + \sigma \right) \right\rbrack
\end{equation}
with $\td s_4^2$ locally K\"ahler-Einstein, $\td s_6^2$ is
Calabi-Yau. There is an $\Lie{SU}(3)$ structure on $\mathcal{M}_6$
that is related to the $\Lie{SU}(2)$ structure on the four-dimensional
base:\footnote{In opposite to \cite{Gauntlett:2004yd}, we include an
  additional overall minus sign in the definition of $J$ in order to
  satisfy \eqref{eq:SU_3-structure_normalization}.

  Also, these forms satisfy a series of useful relations:
  \begin{equation*}
    \begin{aligned}
      \td(e^{\imath\psi^\prime} r^3 \Omega_4) &= 3 \Omega &
      \td\Omega_4 &= 3\imath \sigma\wedge\Omega_4 &
      \td\sigma &= 2 J_4
    \end{aligned}
  \end{equation*}
}
\begin{equation}\label{eq:SU-3-structure}
  \begin{aligned}
    \Omega &= e^{\imath \psi^\prime} r^2 \Omega_4 \wedge \left\lbrack \td r
    + \imath r \left( \frac{\td\psi^\prime}{3} + \sigma \right) \right\rbrack &
    J &= - \left\lbrack r^2 J_4 + r \td r \wedge \left(
      \frac{\td\psi^\prime}{3} + \sigma \right) \right\rbrack
  \end{aligned}
\end{equation}

Specific examples are given by $\mathbb{R}^6$, the singular conifold $T^{1,1}$
and the $Y^{p,q}$ spaces of \cite{Gauntlett:2004yd}. The former two
turn out to be limits of the latter \cite{Gauntlett:2004yd}. Their
metric is given by 
\begin{equation}\label{eq:yPQ_geometries}
  \begin{aligned}
    \td s_6^2 &= r^2 \left\lbrack
        \sqrt{\frac{1-y}{6}} \left(\td\theta^2 + \sin\theta^2
          \td\phi^2\right) + \frac{\td y^2}{\sqrt{w q}} +
        \frac{\sqrt{w q}}{6} \left(\td\beta + \cos\theta\td\phi\right)
      \right. \\ 
    &\qquad\qquad \left. + \left(\frac{1}{3} \td\psi^\prime
          + \sigma\right)^2 \right\rbrack +\td r^2 \\
    \sigma &= \frac{- \cos\theta \td\phi + y \left(
        \td\beta + \cos\theta \td\phi \right)}{3} \\
    w(y) &= \frac{2 (a-y^2)}{1-y} \qquad
    q(y) = \frac{a - 3 y^2 + 2 y^3}{a-y^2}
  \end{aligned}
\end{equation}
while the $\Lie{SU}(2)$ structure is
\begin{equation}
 \begin{aligned}
    \Omega_4 &= \sqrt{\frac{1-y}{6 w q}} (\td\theta + \imath
    \sin\theta\td\phi) \wedge (\td y + \imath \frac{w q}{6} (\td\beta
    + \cos\theta \td\phi) \\
    J_4 &= \frac{1-y}{6} \sin\theta\td\theta\wedge\td\phi +
    \frac{1}{6} \td y \wedge (\td\beta + \cos\beta\td\phi)
  \end{aligned}
\end{equation}
Finally, we introduce a vielbein
\begin{equation}
  \begin{aligned}
    e^1 &= r \sqrt{\frac{1-y}{6}} \td\theta &
    e^2 &= r \sqrt{\frac{1-y}{6}} \sin\theta\td\phi \\
    e^3 &= \frac{r \td y}{\sqrt{wq}} &
    e^4 &= r \frac{\sqrt{wq}}{6} (\td\beta + \cos\theta\td\phi) \\
    e^5 &= r \left( \frac{\td\psi^\prime}{3} + \sigma \right) &
    e^6 &= \td r
  \end{aligned}
\end{equation}

In section \ref{sec:d7-sources-singular} we will use the singular
conifold, since the geometry is less complicated than the more general
metric \eqref{eq:yPQ_geometries}. Here, the internal metric is
\begin{equation}\label{eq:conifold_metric}
  \begin{aligned}
    \td s^2_6 &= \td r^2 + r^2 \left\lbrack \frac{1}{6} \sum_i \left(
        \td\theta_i^2 + \sin\theta_i^2 \td\phi_i^2 \right) 
      + \frac{1}{9} \left( \td\psi^\prime - \sum_i \cos\theta_i
        \td\phi_i \right)^2 \right\rbrack
  \end{aligned}
\end{equation}

\section{Supersymmetry-breaking fluxes}
\label{sec:supersymm-break-flux}

\subsection{The ansatz}
\label{sec:trivial-choice}

There is a difficulty one encounters immediately when solving
\eqref{eq:non-bps-ness_equation}. The right hand side of the equation
takes the form of a complex phase multiplying a real polyform, while
$\Omega$ on the left hand side is intrinsically complex. That is, we have
\begin{equation}
  \begin{aligned}
    \text{l.h.s.} &\in \Omega^*(\mathcal{M}_6)^{\mathbb{C}} &
    \text{r.h.s.} &\in \mathbb{C} \otimes \Omega^*( \mathcal{M}_6 )
  \end{aligned}
\end{equation}
The explicit examples in \cite{Lust:2008zd} are often based on
compactifications on toric manifolds where it is possible to perform
an operation along the lines of separating the real and imaginary
parts of $\Omega$ in a sensible way, yet for the geometries we are
interested that seems not possible.\footnote{One can see this quite
  explicitly by studying the $\Lie{SU}(3)$-structure defined in
  \eqref{eq:SU-3-structure}.
}
Hence, the simplest way to address the
above is to set $\vol_\perp = \vol_6$ and $R = 0$. This gives us the
following equations
\begin{subequations}
\begin{align}
  0 &= \td(e^{3A-\Phi}\Omega) \label{eq:Omega_bps_ness} \\
  H \wedge \Omega &= 4 \imath \mathfrak{r} \vol_6
  \label{eq:non_susy_H}
  \end{align}
\end{subequations}
Hence it is the three-form flux $H$ (and thus $F_3$) that breaks
supersymmetry, while one can expect the BPS-equations of the metric to
remain unchanged. However, the additional flux will deform the
solutions since it appears both in the Einstein equations as well as
the relevant Bianchi identities.

All in all, the NS sector is governed by the BPS equations
\eqref{eq:bps-ness_equations} and \eqref{eq:Omega_bps_ness} as well as
the conditions $H \wedge J = 0$ and $\td H = 0$ on the NS-flux. The RR
fluxes can then be read of from
\eqref{eq:bps-ness_Flux}. \eqref{eq:non_susy_H} can be solved for
$\mathfrak{r}$ and is thus not a condition. Finally one needs to
impose the Bianchi identities \eqref{eq:Bianchi_identity} for suitable
sources. Note that since we set $\theta = 0$, there can be no D5 sources.

Let us for the moment assume that the warp-factor depends only on
$r$. We also take the dilaton to be constant, $\Phi = \Phi_0$. It
follows that $F_1 = 0$. Concerning the NS three-form, we make the ansatz
\begin{equation}\label{eq:kr-flux}
  H = e^{3A-\Phi_0} (\mathfrak{h} \Omega + \bar{\mathfrak{h}}
  \bar{\Omega}) \qquad \mathfrak{h} \in \mathbb{C}, \mathfrak{h} =
  \const
\end{equation}
which corresponds to
\begin{equation}\label{eq:basic_example_susy_breaking}
   \mathfrak{r} = - 2 e^{3A-\Phi_0} \bar{\mathfrak{h}}
\end{equation}
Note that for a fairly typical warp factor this means that the scale
of SUSY-breaking will grow towards the UV, with possibility of
supersymmetry emerging in the IR. Since
$\td(e^{\imath\psi^\prime} r^3 \Omega_4) = 3 e^{3A-\Phi_0} \Omega$, it
follows that the corresponding two-form potential is given by
\begin{equation}
  B = \frac{r^3}{3} e^{-\Phi_0} (\mathfrak{h} e^{\imath\psi^\prime}
  \Omega_4 + \bar{\mathfrak{h}} e^{-\imath\psi^\prime} \bar{\Omega}_4)
\end{equation}
One can also immediately calculate $F_3 = e^{-\Phi} \star_6 H$.
\begin{equation}
  F_3 = - \imath e^{3A-2\Phi_0} (\mathfrak{h} \Omega -
  \bar{\mathfrak{h}} \bar{\Omega})
\end{equation}
The RR three-form satisfies the Bianchi identity, which in the absence
of $F_1$ reduces to closure of the form: $\td F_3 = 0$. The final
step, is to impose
\begin{equation}
  \td F_5 + H \wedge F_3 = 0
\end{equation}
Here we have
\begin{equation}
  \begin{aligned}
    H \wedge F_3 &= -16 e^{6A-3\Phi_0} \mathfrak{h} \bar{\mathfrak{h}}
    \vol_6 \\
    \td F_5 &= \frac{4}{r} e^{2A-\Phi_0} \lbrack r A^{\prime\prime} -
    4 r (A^\prime)^2 + 5 A^\prime \rbrack \vol_6
  \end{aligned}
\end{equation}
The resulting ODE can be solved analytically. Writing the result in
terms of the more conventional warp-factor $A = - \frac{1}{4} \log h$:
\begin{equation}
  \label{eq:r-dependent_warp-factor}
  h = h_0 + \frac{h_1}{r^4} - \frac{4}{3} e^{-2\Phi_0} r^2
    \mathfrak{h} \bar{\mathfrak{h}}
\end{equation}
As we anticipated from \eqref{eq:basic_example_susy_breaking}, the
modifications appear only in the UV. However, they are highly
problematic. At at sufficiently large $r$ $h(r)$ becomes
negative. Since the Ricci scalar behaves as $R \sim h^{-3/2}$, there
is a naked singularity.

To understand things a bit better, it is appropriate to take a look at
the Einstein equation and the energy-momentum tensor
explicitly. Instead of doing so straightaway however, let us
generalize the above discussion, by contrasting the above ansatz for $H$
with supersymmetric flux. In the case of the conifold this
is provided by the Klebanov-Tseytlin background, which can be
generalized to encompass the $Y^{p,q}$ geometries that we discussed in
section \ref{sec:examples}. This was done in
\cite{Herzog:2004tr}. (One can actually obtain Klebanov-Tseytlin
solution as a limit of \cite{Herzog:2004tr}.) So in the following we
work explicitly with \eqref{eq:yPQ_geometries}.

The equations for $H$ are linear, allowing us to use both the
supersymmetry preserving three-form as well as
the SUSY-breaking choice
\begin{equation}\label{eq:full_h_flux}
  \begin{aligned}
    H &= H_{\text{susy}} + H_{\text{nosy}} \\
    H_{\text{susy}} &= (k_1 \td + k_2 \star_6 \td) \left\lbrack \frac{3
      e^{2A} \log r}{2\pi r^2 (1-y)^2} (e^3 \wedge e^4 - e^1 \wedge
    e^2) \right\rbrack
  \end{aligned}
\end{equation}
and $H_{\text{nosy}}$ as above. Since $H_{\text{susy}} \wedge \Omega =
0$, $\mathfrak{r}$ remains unchanged. As we will see, the constants
$k_1$ and $k_2$ enter the background metric only in the form of $k_1^2
+ k_2^2$. $H_{\text{susy}}$ contributes to $F_3$ as
\begin{equation}
  \begin{aligned}
      F_{\text{susy}} &= e^{-\Phi_0} (k_1 \star_6 \td - k_2 \td)  
      \left\lbrack \frac{3 e^{2A} \log r}{2\pi r^2 (1-y)^2} (e^3
        \wedge e^4 - e^1 \wedge e^2) \right\rbrack\\
      &= \frac{3 e^{3A-\Phi_0}}{2\pi r^3 (1-y)^2} (e^1 \wedge e^2 -
      e^3 \wedge e^4) \wedge (e^5 k_1 + e^6 k_2)  
  \end{aligned}
\end{equation}
Writing the warp factor once again in terms of $h(r,y)$, the Bianchi
identity for $F_5$ reduces to the following PDE 
\begin{equation}\label{eq:second_pdf_for_h}
  \begin{aligned}
    0 &= r \partial_r \lbrack 2 e^{2\Phi_0} \pi^2 r^5
    (1-y)^4 \partial_r h \rbrack + \partial_y \lbrack 2 e^{2\Phi_0}
    \pi^2 r^4 (1-y)^4 q w \partial_y h \rbrack \\
    &+ 6 e^{2\Phi_0} \pi^2 r^4 (1-y)^3 q w \partial_y h
    + 32 \pi^2 r^6 (1-y)^4 \mathfrak{h} \bar{\mathfrak{h}} 
    + 9 e^{2\Phi_0} (k_1^2 + k_2^2)
  \end{aligned}
\end{equation}
Following \cite{Herzog:2004tr}, we make the ansatz $h(r,y) = r^{-4}
\lbrack h_r(r) + h_y(y) \rbrack$. This leads to the ODE for $h_r$
\begin{equation}
  \begin{aligned}
    0 &= 16 r^5 \mathfrak{h} \bar{\mathfrak{h}} + e^{2\Phi_0} (r
    h_r^{\prime\prime} - 3 h_r^\prime)
  \end{aligned}
\end{equation}
which is solved by
\begin{equation}\label{eq:h_r_solution}
  \begin{aligned}
    h_r &= h_0 r^4 + h_1 + h_2 \log r
    - \frac{4}{3} e^{-2\Phi_0} \mathfrak{h}\bar{\mathfrak{h}} r^6
  \end{aligned}
\end{equation}
which is identical to
\eqref{eq:r-dependent_warp-factor}. Subsequently, one is left with
another ODE, this time in terms of $y$:
\begin{equation}
  \begin{aligned}
    \partial_y \lbrack 2\pi^2 (1-y) q w \partial_y h_y \rbrack &= \frac{8
    h_2 \pi^2 (1-y)^4 - 9 (k_1^2 + k_2^2)}{(1-y)^3}
  \end{aligned}
\end{equation}
This can be integrated. Since $\partial_y h_y$ is finite at $y_i$ and
$q(y_i) = 0$, the integral of the right hand side has to vanish. This
constraint yields
\begin{equation}
  \begin{aligned}
    h_2 &= \frac{9 (k_1^2 + k_2^2)}{8\pi^2 (1-y_2)^2 (1-y_1)^2}
  \end{aligned}
\end{equation}
fixing $h_2$. The equation for $y$ can then be solved as in
\cite{Herzog:2004tr}. We are mostly interested in the radial behavior
and note that the contribution of the supersymmetric part of the flux
does not allow us to cure the singularity. It is a curious fact that
the different parts of the flux separate so nicely in the PDE for
$h(r,y)$. As soon as we split the equation into two ODEs, the
$H_{\text{susy}}$ is only relevant for $h_y$, while $H_{\text{nosy}}$
appears in the equation for $h_r$.

Having introduced the supersymmetric three-forms, let us take a look
at the Einstein equation and the energy-momentum in Einstein frame,
including possible sources. The technical aspects of the discussion
are of course very similar to \cite{Giddings:2001yu}. From
\eqref{eq:Einstein_equation_Einstein_frame} we find
\begin{equation}
  \begin{aligned}
    0 &= R_{\mu\nu} + \frac{1}{8} g_{\mu\nu} (e^{-\Phi} H^2 + e^\Phi
    F_3^2) - \frac{1}{4} F^5_\mu \cdot F^5_\nu + 2\kappa_{10}^2
    \left\lbrack T^{\text{src}}_{\mu\nu} - \frac{1}{8} g_{\mu\nu}
      (T^{\text{src}})^m_m \right\rbrack
  \end{aligned}
\end{equation}
So, using \eqref{eq:Ricci_tensor_Einstein_frame}, this
becomes\footnote{
Note that the spatial components of the five-form are given by
\begin{equation*}
  F_5 = \td(e^{4A-\Phi}) \wedge \td x^0 \wedge \td x^1 \wedge \td x^2
  \wedge \td x^3
\end{equation*}
which gives (in Einstein frame)
\begin{equation*}
  \begin{aligned}
    F^5_\mu \cdot F^5_\nu &= - \eta_{\mu\nu} e^{4A}
    g^{ij} \partial_i (4A-\Phi) \partial_j (4A-\Phi)
  \end{aligned}
\end{equation*}
}
\begin{equation}
  \begin{aligned}
    e^{4A} \eta_{\mu\nu} \hat{\nabla}^2 A &= \frac{1}{4}
    e^{2A-\frac{3\Phi_0}{2}} \eta_{\mu\nu} H^2 + 4 \eta_{\mu\nu}
    e^{4A} g^{ij} \partial_i A \partial_j A 
    + 2\kappa_{10}^2 T_p \frac{7-p}{16} g_{\mu\nu}
    \frac{\sqrt{-g_{p+1}}}{\sqrt{-g}} \delta(\Sigma)
 \end{aligned}
\end{equation}
Substituting
\begin{equation}
  H^2 = \frac{e^{6A-\frac{\Phi}{2}}}{2} \left\lbrack 9 e^{2\Phi} \frac{k_1^2 +
      k_2^2}{\pi^2 r^6 (1-y)^4} + 32 \mathfrak{h} \bar{\mathfrak{h}} \right\rbrack
\end{equation}
we find\footnote{Since we are not concerned with
  compactifications, we cannot use the argument of
  \cite{Giddings:2001yu} that the right hand side has to vanish.
}
\begin{equation}\label{eq:einstein_pdf}
  \begin{aligned}
    e^{4A} \eta_{\mu\nu} \hat{\nabla}^2 A &= \eta_{\mu\nu} \bigg\{
      \frac{1}{8} e^{8A-2\Phi_0} \left\lbrack 9 e^{2\Phi_0} \frac{k_1^2
          + k_2^2}{\pi^2 r^6 (1-y)^4} + 32 \mathfrak{h}
        \bar{\mathfrak{h}} \right\rbrack  \\
    & + 4 e^{4A} g^{ij} \partial_i A \partial_j
      A \bigg\} + 2\kappa_{10}^2 T_p \frac{7-p}{16} g_{\mu\nu}
    \frac{\sqrt{-g_{p+1}}}{\sqrt{-g}} \delta(\Sigma)
  \end{aligned}
\end{equation}
As one would expect, this is the same PDE \eqref{eq:second_pdf_for_h}
that we previously obtained from the Bianchi identity. Here one sees
very nicely the difference between the SUSY and the non-SUSY
fluxes. The former disappear in the UV and dominate the IR, while the
latter become only relevant in the UV. Finally, note that the
source-term for space-time filling $D_3$ sources is
\begin{equation}
  \begin{aligned}
    \frac{2\kappa_{10}^2 T_3}{4} g_{\mu\nu}
    \frac{\sqrt{-g_{p+1}}}{\sqrt{-g}} \delta^6(y^i) &=
    \frac{2\kappa_{10}^2 T_3}{4} \eta_{\mu\nu}
    e^{8A+\Phi_0} \frac{\delta^6(y^i)}{\sqrt{\hat{g}}} 
  \end{aligned}
\end{equation}
Hence -- as long as the dilaton is constant -- the
three-form fluxes contribute to the energy momentum tensor as a
continuous distribution of D3 branes would.

\subsection{Instead of a no-go theorem}
\label{sec:no_way_out}

One can think of several ways to improve the situation and deal with
the singularity. As we will see -- and as one can check by performing 
explicit calculations -- none of these succeed. Thus, it is tempting
to think that there might be a no-go theorem\footnote{
Of course, one has to be slightly careful here since nonsupersymmetric
deformations have been found that do not encounter the issue of a UV
singularity -- see e.g.~\cite{Kuperstein:2003yt}.}
that does not allow for imaginary self-dual SUSY-breaking three-form flux on
Sasaki-Einstein spaces in the context of supergravity.

There are several simple methods one can think of to change
the PDE \eqref{eq:second_pdf_for_h} that we obtained from the Bianchi
identity. The first of these is changing the relative sign between
$F_3$ and $H$. This Effectively changes the sign in
\eqref{eq:h_r_solution}, leading to a warp factor that (ignoring $y$)
behaves as $r^{-4}$ in the IR and $r^2$ in the UV, again with
supersymmetry breaking in the UV. However, flipping the sign between
the two three-forms is inconsistent. First of all, the Einstein
equation is invariant under this transformation. Yet as we saw,
\eqref{eq:einstein_pdf} is equivalent to \eqref{eq:second_pdf_for_h},
and changing the sign would only modify the latter. Furthermore there
is the issue of the equation of motion for $F_3$ -- i.e.~the Bianchi
identity for $F_7$. The sign here changes as well leading to a further
inconsistency.\footnote{
Since the change of sign leads to an interesting PDE, one can try to
change the sign in both the Bianchi identity and the Einstein
equation. However, the contribution to the latter is of course
positive definite. Hence, there is only the formal possibility of
making the flux imaginary. The substitution $H \mapsto \imath H$, $F_3
\mapsto \imath F_3$ gives indeed a solution to the equations of
motion. As a matter of fact, introducing a phase factor $H \mapsto
e^{\imath \eta} H$, one ends up with a warp factor
\begin{equation*}
  h(r) = h_0 + \frac{h_1}{r^4} - \frac{4}{3} e^{2\imath \eta - 2
    \Phi_0} r^2 \mathfrak{h}\bar{\mathfrak{h}}
\end{equation*}
}

\subsubsection{Additional source terms}
\label{sec:addit-source-terms}

Sticking with more physical possibilities, one can look into
modifying the Bianchi identity and Einstein equation by the addition
of backreacting sources. The simplest of course are space-time filling
objects with codimension 6. That is 3-branes and O3-planes. From the
Bianchi identity \eqref{eq:Bianchi_identity} it follows that the
appropriate source term is $j_6$. One has to be careful with
the sign here. As discussed in the appendix \ref{sec:type-iib},
3-branes couple to the four-form potential via $T_3 \int C_4 \wedge
\sigma(j_6)$, yet $\sigma(j_6) = -j_6$. We can write a generic source as
\begin{equation}
  j_6 = - e^{-3\Phi_0} \rho \vol_6  
\end{equation}
for a generic density function $\rho : \mathcal{M}_6 \to
\mathbb{R}$. Once can deal with the $y$-dependence as 
in section \ref{sec:trivial-choice}, so for simplicity we assume that
$h$ depends on $r$ alone. The resulting ODE is
\begin{equation}
  0 = r (16 \mathfrak{h} \bar{\mathfrak{h}} + \rho) + e^{2\Phi_0} (5
  h^\prime + r h^{\prime\prime})
\end{equation}
It becomes clear that only a negative, constant $\rho$ allows to
cancel the three-form flux charge. This corresponds to a
constant background charge density extending all the way to the far
UV, a rather problematic assumption. Since this analysis
concerns only the Bianchi identity, the question remains whether the
additional charge is that of anti D3-branes or of orientifold
planes. Here, the Einstein equation \eqref{eq:einstein_pdf} confirms
that we are actually dealing with negative tension objects.

For completeness, let us look into possible radial dependence of
$\rho$. In this case the new branes are smeared over the
Sasaki-Einstein base. The simplest example $\rho(r) = \delta(r-r_1)$
changes the warp factor by a further negative term:
\begin{equation}
  h(r) \mapsto h(r) - e^{-2\Phi_0} \frac{r_1}{4} \left(1 -
    \frac{r_1^4}{r^4}\right) \theta(r - r_1)
\end{equation}
In light of the fact that the relevant PDEs are all linear in $h$,
this result should not be surprising.

Before we move on to D7-branes, let us briefly consider another
possibility: Since we have $h \to 0$ at the singularity, it might be
possible to reconsider the solution as a compact one. I.e.~for $h$ as
in \eqref{eq:r-dependent_warp-factor} with $h_0 = 0$ the
warp-factor vanishes at $r_0 = \left( \frac{3 h_1}{4
    \mathfrak{h}\bar{\mathfrak{h}}} \right)^{1/6}$ and thus the volume
of the internal manifold goes to zero. If one considers the geometry
as defined on the compact interval $r \in \lbrack 0, r_0 \rbrack$,
with a suitable negative tension source -- that we found in the last
paragraph -- at $r_0$ to cancel tadpoles. Yet by analyzing the metric
around $r_0$, we find that the singularity is not conical.\footnote{To
  understand this in principle, let us take the warp factor to be of
  the form $h = r^{-4} - r^2$. If one introduces a new coordinate via
  $r = 1 - \frac{\rho^4}{6}$, one finds
\begin{equation*}
  \sqrt{h} (\td r^2 + r^2 \td s_5^2) = \rho^2 \td s_5^2 - \frac{5}{24}
  \rho^6 \td s_5^2 + \frac{4}{9} \rho^8 \td\rho^2 + \mathcal{O}(\rho^9)
\end{equation*}
That is, it is not possible to locally write the metric as $\td \rho^2
+ \rho^2 \td s_5^2$ and one can only satisfy the relevant junction
conditions by assuming that there is a \emph{thick} object with
negative tension at $r_0$; again, not a physically viable assumption.
}

\subsubsection{D7 sources on the singular conifold}
\label{sec:d7-sources-singular}

So finally we turn to the inclusion of D7-branes, which corresponds to
a non-constant dilaton in the background. The general idea is that the
orientation of the branes and the three-form flux is such that the
$H$-field induces negative D3-charge onto the branes, canceling or at
least softening the effect of the SUSY-breaking terms in the Bianchi
identity for the five-form. Furthermore the inclusion of such branes
will induce a non-trivial profile for the dilaton, which might also
help matters. The inclusion of backreacting D7-branes in
Sasaki-Einstein geometries was first studied in
\cite{Benini:2006hh}. See also \cite{Bigazzi:2008zt} and
\cite{Nunez:2010sf}. In general the D7-branes are embedded in such a
way that the $\Lie{U}(1)$-fiber in the five-dimensional
Sasaki-Einstein space is deformed. I.e.~we need to make an ansatz for
the metric \eqref{eq:conifold_metric} allowing for such deformations
and introduce here the relevant vielbein:
\begin{equation}\label{eq:conifold_vielbein}
  \begin{aligned}
    e^1 &= \frac{e^g}{\sqrt{6}} \td\theta_1 &
    e^2 &= \frac{e^g}{\sqrt{6}} \sin\theta_1 \td\phi_1 \\
    e^3 &= \frac{e^g}{\sqrt{6}} \td\theta_2 &
    e^4 &= \frac{e^g}{\sqrt{6}} \sin\theta_2 \td\phi_2 \\
    e^5 &= e^f \td r &
    e^6 &= \frac{e^f}{3} (\td\psi^\prime - \sum_i \cos\theta_i \td\phi_i)
  \end{aligned}
\end{equation}
Note that in opposite to the above references we also include the
warp factor for the internal space -- $e^{-2A}$ -- in the
ten-dimensional geometry and choose a different radial coordinate. The
$\Lie{SU}(3)$ structure is given by
\begin{equation}
  \begin{aligned}
    J &= e^1 \wedge e^2 + e^3 \wedge e^4 + e^5 \wedge e^6 \\
    \Omega &= e^{\imath \psi^\prime} (e^1 + \imath e^2) \wedge (e^3 +
    \imath e^4) \wedge (e^5 + \imath e^6)
  \end{aligned}
\end{equation}
While the SUSY-breaking flux remains as in \eqref{eq:kr-flux} (except
that the dilaton is no longer constant), we generalize the ansatz for
the SUSY flux slightly to allow for different radial dependence.
\begin{equation}\label{eq:cfd_susy_flux}
  \begin{aligned}
    B_{\text{susy}} &= k_1(r) (\sin\theta_1 \td\theta_1 \wedge
    \td\phi_1 - \sin\theta_2 \td\theta_2 \wedge \td\phi_2) \\
    &+ \frac{k_2}{3} (\cos\theta_1 \cos\theta_2 \td\phi_1 \wedge
    \td\phi_2 - \cos\theta_1 \td\phi_1 \wedge \td\psi + \cos\theta_2
    \td\phi_2 \wedge\psi)
  \end{aligned}
\end{equation}

Proceeding as usual, one finds BPS-like-equations
\begin{equation}
  \begin{aligned}
    f^\prime &= 3 - 2 e^{2f-2g} &
    g^\prime &= e^{2f-2g} + \frac{\Phi^\prime}{2}
  \end{aligned}
\end{equation}
When it comes to imposing the Bianchi identities, things are a bit
more complicated due to the D7s. Also, instead of using the forms
$j_i$ appearing in \eqref{eq:Bianchi_identity} it is actually simpler
to use the $\Theta_i$ defined in \eqref{eq:theta_sources}. The reason
is that the $j_i$ do not just correspond to sources alone, but also to
induced charge, as one can see from the form of the Wess-Zumino term
when written in terms of $j$ -- see \eqref{eq:source-term-action}.

Again, there are no five-brane source terms and therefore we need to
impose $\Theta_4 = 0$. We have
\begin{equation}
  \begin{aligned}
    \Theta_4 &= 2 \sqrt{6} e^{4A-f-3g-\Phi} k_2 \Phi^\prime
    (\cot\theta_2 e^{1246} - \cot\theta_1 e^{2346}) \\
    &+ \frac{2}{3} e^{4A-f-2\Phi} \lbrack \re(\mathfrak{h}
    e^{\imath\psi^\prime}) e^{1356} +
    \im(\mathfrak{h}e^{\imath\psi^\prime}) (e^{1456} + e^{2356})
    \rbrack \lbrack (\Phi^\prime)^2 - \Phi^{\prime\prime}) \\
    &+ \frac{2}{3} e^{4A-2(f+g+\Phi)} \lbrack 3 e^\Phi k_2
    \cot\theta_1 \cot\theta_2 + e^{f+2g} \re
    (\mathfrak{h}e^{\imath\psi^\prime}) \rbrack e^{2456}
    (\Phi^\prime)^2 - \Phi^{\prime\prime}) \\
    &+ 6 e^{4A-2(f+g)-\Phi} (e^{1256} - e^{3456}) \{ 2 k_1^\prime
    \Phi^\prime - k_1^{\prime\prime} + k_1 \lbrack \Phi^{\prime\prime}
    - (\Phi^\prime)^2 \rbrack \}
  \end{aligned}
\end{equation}
For this to vanish, we have to impose
\begin{equation}
  \begin{aligned}
    \Phi &= \Phi_0 - \log(r+r_0) &
    k_1 &= - \frac{\tilde{k}_1}{r+r_0} &
    k_2 &= 0
  \end{aligned}
\end{equation}
In contrast to the previous case, there is no longer equivalence
between $H$ and its Hodge dual, which is of course due to the fact
now the Bianchi identity for $F_3$ is twisted by $F_1$. An
interesting difference to the supersymmetric case is that we are not
free to choose the dilaton. One can immediately calculate
\begin{equation}
  \begin{aligned}
    F_1 &= \frac{e^{-\Phi_0}}{3} ( \td\psi^\prime - \cos\theta_1
    \td\phi_1 - \cos\theta_2 \td\phi_2 ) \\
    \Theta_2 &= \frac{1}{3} e^{-\Phi_0} (\sin\theta_1 \td\theta_1 \wedge
    \td\phi_1 + \sin\theta_2 \td\theta_2 \wedge \td\phi_2 )
  \end{aligned}
\end{equation}
It follows that the additional D7 sources correspond to massless
flavors in the gauge theory. This is in stark contrast to the
supersymmetric case, where there are solutions dual to massive
flavors -- see \cite{Bigazzi:2008zt}. However, one can see directly
from the form of $\Theta_4$ that setting $\mathfrak{h} \to 0$ removes
any condition on the dilaton.

Since the dilaton is no longer arbitrary, the BPS-like equations for
$f$ and $g$ can be solved.
\begin{equation}
  \begin{aligned}
    f &= e^{2 c_f}{3} + 3 r + r_0 - \frac{1}{3} \log \left\lbrack
      54 c_g + e^{6r} \left( r + r_0 - \frac{1}{6} \right)
    \right\rbrack \\
    g &= \frac{1}{6} \left\{ 4 c_g + 6 r_0 - 3 \log (r + r_0) + \log \left\lbrack 54
        c_g + e^{6r} \left( r + r_0 - \frac{1}{6} \right)
      \right\rbrack \right\}
  \end{aligned}
\end{equation}
Finally, we have to impose $\Theta_6 = 0$. This gives the following ODE:
\begin{equation}\label{eq:big_bad_ODE__wolf}
  \begin{aligned}
    0 &= 4 e^{4A} \lbrack 2 e^{2f+4g} (r + r_0)^6 \mathfrak{h}
    \bar{\mathfrak{h}} + 9 e^{2\Phi_0} \tilde{k}_1 \rbrack 
    + e^{2(g + \Phi_0)} (r + r_0)^2 \{ - 2 e^{2f} (r + r_0) \lbrack 1
    + 4 A^\prime (r + r_0) \rbrack \\
    &+ e^{2g} \lbrack 1 + 2 (r + r_0) (2 A^\prime + 4 (r + r_0)
    (A^{\prime})^2 - (r + r_0) A^{\prime\prime} ) \rbrack \}
  \end{aligned}
\end{equation}

In order to discuss \eqref{eq:big_bad_ODE__wolf}, let us impose impose
$c_g = c_f = 0$ and $r_0 = \frac{1}{6}$. $c_f$ and $c_g$
are simply scales for the two internal two-spheres while $r_0$ can be
expected to set the minimum for the radial coordinate. Since we are
mainly interested in the question whether it might be possible to cure
the UV singularity, this should be a good assumption. Similarly,
setting the supersymmetric part of the three-form flux to zero is also
reasonable. With these assumptions, let us first discuss the SUSY
solutions to \eqref{eq:big_bad_ODE__wolf} -- i.e.~we set $\mathfrak{h}
\to 0$. Then working in terms of the warp factor $h(r)$, we find
\begin{equation}\label{eq:big_bad_wolf__SUSY_solution}
  \begin{aligned}
    h(r) &= (6r+1) (h_0 + h_1 \Gamma_{\frac{1}{3}}(4r) 
    + \frac{36 \times 2^{1/3} r^{2/3} \tilde{k}_1^2 \lbrack 3 \times
      2^{2/3} r^{1/3} - 2 e^{4r} (6r+1) \Gamma_{\frac{1}{3}}(4r) \rbrack}{(e^{6r+1}r)^{2/3}}
  \end{aligned}
\end{equation}
with the incomplete gamma function
\begin{equation}
  \Gamma_a(z) = \int_z^\infty t^{a-1} e^{-t} \td t 
\end{equation}
One can check explicitly that this leads to a well-defined metric. That
is, as long as $\tilde{k}_1$ is not too large, we have $h > 0$.

Turning to the non-SUSY case, we take set $\tilde{k}_1$ to zero as
well to keep things manageable. That should be reasonable since we
have already seen previously that the additional SUSY flux does not
cure the singularity. One finds
\begin{equation}
  \begin{aligned}
    h &= \frac{6r+1}{18} \left\{ 18 h_0 + 9 \times 2^{1/3} h_1
      \Gamma_{\frac{1}{3}}(4r) + 2^{5/3} (-1)^{1/3} \mathfrak{h}
      \bar{\mathfrak{h}} \Gamma_{\frac{4}{3}}(-2r) \right\}
  \end{aligned}
\end{equation}
Here, both the factor $(-1)^{1/3}$ and the gamma function with
negative argument are complex, showing that it is not possible to add
D7 branes while breaking supersymmetry.

\subsubsection{A different choice for the three-form flux}
\label{sec:diff-choice-three}

Let us finally turn to what is probably the most obvious question --
whether there is a different three-form that one can use for the
SUSY-breaking flux instead of the obvious choice we used so far,
equation \eqref{eq:kr-flux}.

The problem is that the flux has not only to satisfy the algebraic
relation $H \wedge J = 0$, but has also to satisfy the various Bianchi
identities. Not least of all it has to be well-defined. Let us give
two examples.

In the case of the conifold, there is
\begin{equation}
  \frac{e^{f+2g}}{6} \sin\theta_1 \sin\theta_2 \td r \wedge
  \td\theta_1 \wedge \td\theta_2 
\end{equation}
yet here it is not possible to impose the Bianchi identity for
$F_3$. In the case of $\AdS_5 \times S^5$ there are several possible
ans\"atze, since the internal geometry is $\mathbb{R}^6$. As an
example, let us choose
\begin{equation}
  B = b(y_1) \td y_3 \wedge \td y_5 
\end{equation}
where we have introduced cartesian coordinates $y_i$ on
$\mathbb{R}^6$. The Bianchi identity for $F_3$ leads to the linear
relation
\begin{equation}
  b(y_1) = b_1 y_1 
\end{equation}
while we find for the warp-factor ($y^2 = \sum_i y_i^2$)
\begin{equation}
  A(y) = - \frac{1}{4} \log \left( h_0 + \frac{h_0}{y^4} - \frac{b_1^2}{12}
    y^2 \right)
\end{equation}
Which has the same UV singularity as
\eqref{eq:r-dependent_warp-factor}. This is a reasonable result, since
in both cases we add constant flux to the background geometry.

In general, if the form $J$ is given by
  \begin{equation}
    J = \sum_{i=1}^3 e^i \wedge e^{i+3}
  \end{equation}
  the algebraic relation is satisfied by the eight three-forms
  \begin{equation}
    \begin{aligned}
      e^i \wedge e^j \wedge e^k \qquad i \in \{1,4\}, j \in \{2,5\}, k
      \in \{3,6\}
    \end{aligned}
  \end{equation}
  which are in general often ill-defined though and certainly not
  necessarily closed. This is not the only class of three-forms
  satisfying the relation as can be understood from the
  Klebanov-Tseytlin case. Here, one makes use of the fact that there
  are two non-trivial two-cycles in the K\"ahler-Einstein base (see
  \cite{Gauntlett:2004yd}). The ansatz for the flux is then based on a
  linear combination of the two and satisfies $H \wedge J = 0$ because
  the contribution from each form cancels that of the other. This is
  straightforward to see in the case of the conifold -- see the first
  line in our ansatz for the supersymmetric flux there, equation
  \eqref{eq:cfd_susy_flux}. If one tries to make this sort of ansatz
  for the SUSY-breaking case, one always finds $H \wedge \Omega = 0$.

\section{Conclusions}
\label{sec:conclusions}

In the main part of this paper, we attempted a rather exhaustive study
of nonsupersymmetric imaginary self-dual three-form flux on
Sasaki-Einstein backgrounds. While we were able to find analytic
solutions, all of these were plagued by problematic issues, most
notably a naked curvature singularity in the UV that raises the
question whether there is a well-defined UV completion. Since all our
attempts at curing the singularity failed one might conjecture that it
is not possible to do so within the context of
supergravity. Comparisons with the supersymmetric flux suggest that
this is due to the fact that the flux does not disappear in the UV,
yet the radial behavior of the flux is fully determined by the Bianchi
identities on $H$ and $F_3$.

So far, we have tried to remove the singularity by modifying the
differential equations resulting from the Bianchi identity and
Einstein equations -- without success. An alternative is to try to
push the singularity to infinity by rescaling the solution in a
suitable fashion. From a first study it appears, that when doing so one usually
ends up with a supersymmetric solution without three-form flux, yet
this question is not fully answered.

Of course, not all singularities in supergravity backgrounds are
problematic in the context of string theory, and therefore it might be
possible to find a more suitable UV completion for the backgrounds
discussed. Yet in the context of gauge/string duality it would be
certainly more desirable to study solutions that restore SUSY in the
UV and break it in the IR instead of the opposite.

One has to wonder whether it might be possible to improve the
situation by finding a way to modulate the behavior of the flux for
large $r$. While this does not seem possible within the DWSB
framework, one might be able to combine the ans\"atze presented here
with the alternative approaches mentioned in the
introduction. E.g.~even maintaining the ansatz \eqref{eq:kr-flux} with
$\Phi \neq \Phi_0$ one should obtain a limited amount of total flux as
long as the dilaton grows sufficiently fast towards the UV.

Let us finish with the remark that both the DWSB approach as well as
the further restrictions we have imposed here limit the possible
solutions considerably. Hence to make progress with the idea of
maintaining the existence of a $G$-structure while breaking
supersymmetry, it should be instructive to study the $G$-structure and
the violation of the associated BPS-equations in the case of known
non-supersymmetric deformations such as \cite{Aharony:2002vp} and
\cite{Kuperstein:2003yt}.

\begin{acknowledgements}
I would like to thank Shinji Mukohyama, Carlos N\'u\~nez, Alfonso
Ramallo, Shigeki Sugimoto and Taizan Watari for valueable discussions as well as
comments on the manuscript. Further thanks are due to Carlos
N\'u\~nez, who was involved in the initial stages of this project. I
am supported by a Postdoctoral Fellowship of the Japan Society for the
Promotion of Science (JSPS). This work was
supported by World Premier International Research Center Initiative
(WPI Initiative), MEXT, Japan.
\end{acknowledgements}

\appendix

\section{Conventions}
\label{sec:conventions}

\subsection{Differential geometry}
\label{sec:diff-geom}

We use the same conventions as in \cite{Lust:2008zd} and
\cite{Koerber:2007hd}. The ten- and six-dimensional Hodge duals are
defined as follows
\begin{equation}
  \label{eq:Hodge_duals}
  \begin{aligned}
    \star_6 \omega &= \frac{\sqrt{g_6}}{p! (6-p)!} \epsilon_{m_1 \dots
      m_6} \omega^{m_{7-p} \dots m_6} \td y^{m_1} \wedge \dots \wedge
    \td y^{m_{6-p}} \\
    \star_{10} \omega &= - \frac{\sqrt{-g_{10}}}{p! (10-p)!}
    \epsilon_{M_1 \dots M_{10}} \epsilon_{M_1 \dots M_{10}}
    \omega^{M_{11-p} \dots M_{10}} \td x^{M_1} \wedge \dots \wedge \td
    x^{M_{10-p}}
  \end{aligned}
\end{equation}
As a consequence, we have
\begin{equation}
  \begin{aligned}
    \eta \wedge \star_6 \lambda &= (-1)^{p(6-p)} \frac{1}{p!} \eta_{m_1 \dots m_p}
    \lambda^{m_1 \dots m_p} \vol_6
  \end{aligned}
\end{equation}
$\td_H$ is the twisted exterior derivative and
acts on forms as
\begin{equation}
  \label{eq:twisted_exterior_derivative}
  \td_H \omega = (\td + H \wedge) \omega
\end{equation}
There is also a modified hodge dual $\tilde{\star}_6$, that includes
the action of the operator $\sigma$:
\begin{equation}
  \tilde{\star}_6 \equiv \star_6 \circ \sigma 
\end{equation}
where $\sigma$ reverses all indices of a $p$-form
\begin{equation}\label{eq:sigma_operator}
  \sigma (\omega) = \frac{1}{p!} \omega_{M_p \dots M_1} \td x^{M_1}
  \wedge \dots \wedge \td x^{M_p} = (-1)^{\frac{(p-1)p}{2}} \omega 
\end{equation}
The forms $J$ and $\Omega$ defining the $\Lie{SU}(3)$ structure
satisfy the following relations
\begin{equation}\label{eq:SU_3-structure_normalization}
  \begin{aligned}
    \frac{1}{3!} J \wedge J \wedge J &= -\frac{\imath}{8} \Omega
    \wedge \bar{\Omega} = \vol_6 & &\text{and} &
    \star_6 \Omega &= -\imath \Omega
  \end{aligned}
\end{equation}
For an $n$-dimensional internal space (in our case, $n = 6$), the
Mukhai pairing and its generalization are defined as follows
\begin{equation}
  \begin{aligned}
    \langle \omega, \chi \rangle &= \omega \wedge \sigma(\chi) \vert_n
    &
    \langle \omega, \chi \rangle_k &= \omega \wedge \sigma(\chi) \vert_k
  \end{aligned}
\end{equation}

\subsection{Type IIB supergravity}
\label{sec:type-iib}

The source-terms $j$ arise from the action of space-time filling
branes, described by
\begin{equation}\label{eq:source-term-action}
  \begin{aligned}
    S_{\text{src}} &= - T_p \int_{\mathcal{M}_{10}} \Psi \wedge \sigma
    (j) + T_p \int_{\mathcal{M}_{10}} \frac{C + \tilde{C}}{2} \wedge
    \sigma(j) \\
    &= - T_p \int_\Sigma e^{-\Phi} \sqrt{ - g_{\text{ind}} +
      \mathcal{F}} + T_p \int_\Sigma \frac{C + \tilde{C}}{2} \wedge e^{\mathcal{F}}
  \end{aligned}
\end{equation}
where $\Sigma$ is the branes' world volume. The calibration form
$\Psi$ is\footnote{With
  \begin{equation*}
    \vol_{1,3} = e^{4A} \td x^0 \wedge \td x^1 \wedge \td x^2 \wedge
    \td x^3
  \end{equation*}
}
\begin{equation}
  \begin{aligned}
    \Psi &= e^{-\Phi} \vol_{1,3} \wedge \re (e^{\imath \theta} e^{\imath J})
  \end{aligned}
\end{equation}
The $\tilde{C}$ appearing in \eqref{eq:source-term-action} are the
dual potentials. I.e.~the RR field strengths are related to their
potentials via
\begin{equation}
  \begin{aligned}
    F &= \td_H C = \star_{10} \td_H \tilde{C}
  \end{aligned}
\end{equation}
There is a subtlety when matching the components of the polyform $j$
with possible sources, since $j$ also accounts for induced charges. In
other words, the presence of say a D7-brane can induce a D3 charge,
leading to $j_6 \neq 0$ in the absence of D3 sources. To account for
this, one can introduce a further polyform $\Theta$ and write the
Wess-Zumino term as
\begin{equation}
  \begin{aligned}
    T_p \int_{\mathcal{M}_{10}} \frac{C + \tilde{C}}{2} \wedge
    \sigma(j) &=
    T_p \int_{\mathcal{M}_{10}} \frac{C + \tilde{C}}{2} \wedge
    e^{\mathcal{F}} \wedge \Theta
  \end{aligned}
\end{equation}
Matching terms on both sides, one finds
\begin{equation}\label{eq:theta_sources}
  \begin{aligned}
    j_2 &= -\Theta_2 \\
    j_4 &= \Theta_4 + B \wedge \Theta_2 \\
    j_6 &= - (\Theta_6 + B \wedge \Theta_4 + \frac{1}{2} B^2 \wedge \Theta_2) \\
    j_8 &= \Theta_8 + B \wedge \Theta_6 + \frac{1}{2} B^2 \wedge
    \Theta_4 + \frac{1}{3!} B^3 \wedge \Theta_2
  \end{aligned}
\end{equation}

We will use the Einstein frame\footnote{
Recall that $(g_s)_{MN} = e^{\frac{\Phi}{2}} (g_E)_{MN}$. Furthermore,
under Weyl transformations of the metric, the Ricci scalar behaves as
\begin{equation*}
   R\lbrack e^{\alpha \Phi} g\rbrack = e^{-\alpha\Phi} \left\{ R\lbrack
   g\rbrack - (D-1)\alpha \nabla^2 \Phi - \frac{(D-1)(D-2)}{4}
   \alpha^2 \partial\Phi^2 \right\}
\end{equation*}
}
only when studying the energy-momentum
tensor. Thus the use of the democratic formalism is rather
cumbersome. Hence we drop $F_7$ and $F_9$ when working in Einstein frame:
\begin{equation}
  \begin{aligned}
    S_E = \frac{1}{2\kappa_{10}^2} \int \td^{10} x \sqrt{-g}
    \bigg\lbrack &R - \frac{1}{2} \partial\Phi^2 -
      \frac{e^{-\Phi}}{2} H^2 \\
      &- \frac{1}{2} \bigg( e^{2\Phi} F_1^2 +
        e^\Phi F_3^2 + \frac{1}{2} F_5^2 \bigg) \bigg\rbrack
  \end{aligned}
\end{equation}
This leads to an equation of motion
\begin{equation}\label{eq:Einstein_equation_Einstein_frame}
  \begin{aligned}
    0 &= R_{mn} - \frac{1}{2} \partial_m \Phi \partial_n \Phi -
    e^{-\Phi}{2} H_m H_n + \frac{1}{8} g_{mn} (e^{-\Phi} H^2 + e^\Phi
    F_3^2) \\
    &- \frac{1}{2} \left( e^{2\Phi} F^1_m F^1_n + e^\Phi F^3_m \cdot
      F^3_n + \frac{1}{2} F^5_m \cdot F^5_n \right) 
    + 2\kappa_{10}^2 \left( T^{\text{src}}_{mn} - \frac{1}{8} g_{mn}
      g^{kl} T^{\text{src}}_{kl} \right) \\
    T^{\text{src}}_{mn} &= \frac{1}{\sqrt{-g}} \frac{\delta}{\delta
      g^{mn}} S_{\text{src}}
  \end{aligned}
\end{equation}

Furthermore, the Ricci tensor and scalar of the Einstein-frame metric
\begin{equation}
  \td s^2 = e^{2A-\frac{\Phi}{2}} \td x^\mu \td x_\mu +
  e^{-2A-\frac{\Phi}{2}} \hat{g}_{ij} \td y^i \td y^j
\end{equation}
are given by
\begin{equation}\label{eq:Ricci_tensor_Einstein_frame}
  \begin{aligned}
    R_{\mu\nu} &= e^{4A} \eta_{\mu\nu} \left\lbrack - \frac{1}{2}
      \hat{\nabla}^2 \left( 2A - \frac{\Phi}{2} \right) + \partial^i
      \Phi \partial_i \left( 2A - \frac{\Phi}{2} \right) \right\rbrack
    \\
    &= -\frac{1}{2} e^{4A} \eta_{\mu\nu} \left\lbrack
      e^{-2A+\frac{\Phi}{2}} \hat{\nabla}^2 e^{2A-\frac{\Phi}{2}}
      + \partial\left( 2A+\frac{\Phi}{2}\right)^2 - \partial\Phi^2 \right\rbrack \\
    R_{ij} &= \hat{R}_{ij} + 2 \hat{\nabla}_i \partial_j \Phi -
    8 \partial_i A \partial_j A + 2 (\partial_i A \partial_j \Phi
    + \partial_i \Phi \partial_j A) + \frac{1}{2} \partial_i
    \Phi \partial_j \Phi \\
    &+ \hat{g}_{ij} \left\lbrack \frac{1}{2} \hat{\nabla}^2
      \left(2A+\frac{\Phi}{2}\right) - \partial^l \Phi \partial_l
      \left(2A+\frac{\Phi}{2}\right) \right\rbrack \\
    R &= e^{2A+\frac{\Phi}{2}} \left( \hat{R} + 2 \hat{\nabla}^2 A -
      8 \partial A^2 + \frac{9}{2} \hat{\nabla}^2 \Phi -
      \frac{9}{2} \partial \Phi^2 \right)
  \end{aligned}
\end{equation}
with $R_{\mu i} = R_{i \mu} = 0$. The $(\mu, \nu)$ components of the
Einstein tensor are
\begin{equation}\label{eq:Einstein_tensor}
  \begin{aligned}
    R_{\mu\nu} - \frac{1}{2} g_{\mu\nu} R &= e^{4A} \eta_{\mu\nu}
    \left\lbrack -2 \hat{\nabla}^2 (A+\Phi) + 4 \partial A^2 +
      \frac{7}{4} \partial\Phi^2 + 2 \partial^i \Phi \partial_i A -
      \frac{1}{2} \hat{R} \right\rbrack
  \end{aligned}
\end{equation}
We used that
\begin{equation}
  \begin{aligned}
    \Gamma^i_{\mu\nu} &= - \frac{1}{2} e^{4A} \eta_{\mu\nu}
    \hat{\partial}^j \left(2A-\frac{\Phi}{2}\right) \qquad
    \Gamma^\lambda_{\mu i} = \frac{1}{2} \delta^\lambda_\mu \partial_i
    \left(2A-\frac{\Phi}{2}\right) \\
    \Gamma^l_{ij} &= \hat{\Gamma}^l_{ij} + \frac{1}{2} \left\lbrack (
      \delta^l_j \partial_i + \delta^l_i \partial_j - \hat{g}_{ij}
      \hat{\partial}^l) \left(-2A-\frac{\Phi}{2} \right) \right\rbrack
  \end{aligned}
\end{equation}

\section{The full set of DWSB equations}
\label{sec:full-set-dwsb}

We list the DWSB equations that were derived in \cite{Lust:2008zd} in
terms of the pure spinors $\Psi_{1,2}$. In the special case of type
IIB $\Lie{SU}(3)$-structure backgrounds, these take the form
\begin{equation}
  \begin{aligned}
    \Psi_1 &= e^{\imath \theta} e^{\imath J} &
    \Psi_2 &= e^{-\imath \theta} \Omega
  \end{aligned}
\end{equation}
We begin with those equations
that remain the same as in the supersymmetric case
\begin{equation}\label{eq:dwsb_list_bps_ness}
  \begin{aligned}
    \td_H (e^{4A-\Phi} \re\Psi_1) &= e^{4A}\tilde{F} &
    \td_H (e^{2A-\Phi} \im\Psi_1) &= 0
  \end{aligned}
\end{equation}
The equation for $\Psi_2$ however is modified from its supersymmetric
form to
\begin{equation}\label{eq:dwsb_list_non_bps-ness}
  \td_H (e^{3A-\Phi} \Psi_2) = 4 \imath \mathfrak{r} (-1)^{\vert \Psi_2 \vert}
  e^{3A-\Phi} \frac{\sqrt{\det g\vert_\Pi}}{\sqrt{\det (g\vert_\Pi + R)}}
  e^{-R} \wedge \sigma(\vol_\perp)
\end{equation}
As discussed in the main text, $\mathfrak{r}:\mathcal{M}_6 \to
\mathbb{C}$
parametrizes the supersymmetry-breaking, the tangent space
$T_*\mathcal{M}_6$ can be decomposed as $T_*\mathcal{M}_6 = T_*\Pi
\oplus T_*\Pi^\perp$ and $R \in \bigwedge T^*\Pi$. Recalling that the
degree of a polyform is the form-degree of its highest-degree form, we
note that 
\begin{equation}
  \vert \Psi \vert = \deg(\Psi) \mod 2
\end{equation}
The Bianchi identities remain $\td H = 0$ and $\td_H F = -j$. In the
supersymmetric case ($\mathfrak{r} = 0$) any solution to the above
equations also solves the full equations of motion. However, for
$\mathfrak{r} \neq 0$, one also has to impose
\begin{equation}\label{eq:b-field_eq}
  \begin{aligned}
    \td \lbrack e^{4A-\Phi} \langle \im(\mathfrak{r}^* \Psi_2), 3
    \re\Psi_1 + \frac{1}{2} (-1)^{\vert\Psi_2 \vert} \Lambda^{mn}
    \gamma_m \re\Psi_1 \gamma_n\rangle_3\rbrack &= 0
  \end{aligned}
\end{equation}
and
\begin{equation}\label{eq:einstein_eq}
  \begin{aligned}
    \im \{ \langle g_{k(m} \td y^k \wedge \imath_{n)}, \td_H \lbrack
    e^{A-\Phi} \mathfrak{r}^* (3 \re \Psi_1 + \frac{1}{2}
    (-1)^{\vert\Psi_2\vert} \Lambda^{kr} \gamma_k \re\Psi_1 \gamma_r )
    \rbrack \rangle \} &= 0
  \end{aligned}
\end{equation}
with
\begin{equation}
  \begin{aligned}
    \Lambda &= 1_\perp - (g\vert_\Pi + R)^{-1}(g\vert_\Pi - R)
  \end{aligned}
\end{equation}
and
\begin{equation}
  \begin{aligned}
    \gamma_m \omega &= (\imath_m + g_{mn} \td y^n \wedge) \omega &
    \omega \gamma_m &= (-1)^{\vert\omega\vert+1} (\imath_m - g_{mn}
    \td y^n \wedge) \omega
  \end{aligned}
\end{equation}
For $R = 0$, the above simplify considerably due to the identity
\begin{equation}
  \begin{aligned}
    3 \re\Psi_1 - \frac{1}{2} \Lambda^{mn} \gamma_m \re\Psi_1 \gamma_n
    &= 4 (\re\Psi_1 - \vol_\Pi)
  \end{aligned}
\end{equation}



\begin{thebibliography}{99}


\bibitem{Witten:1998zw}
 E.~Witten,
 Adv.\ Theor.\ Math.\ Phys.\  {\bf 2}, 505 (1998)
 [arXiv:hep-th/9803131].

\bibitem{Sakai:2004cn}
 T.~Sakai and S.~Sugimoto,
 Prog.\ Theor.\ Phys.\  {\bf 113}, 843 (2005)
 [arXiv:hep-th/0412141].

\bibitem{Khavaev:1998fb}
 A.~Khavaev, K.~Pilch and N.~P.~Warner,
 Phys.\ Lett.\  B {\bf 487}, 14 (2000)
 [arXiv:hep-th/9812035].

\bibitem{Gubser:1999pk}
 S.~S.~Gubser,
 arXiv:hep-th/9902155.

\bibitem{Aharony:2002vp}
 O.~Aharony, E.~Schreiber and J.~Sonnenschein,
 JHEP {\bf 0204}, 011 (2002)
 [arXiv:hep-th/0201224].

\bibitem{Borokhov:2002fm}
 V.~Borokhov and S.~S.~Gubser,
 JHEP {\bf 0305}, 034 (2003)
 [arXiv:hep-th/0206098].

\bibitem{Kuperstein:2003yt}
 S.~Kuperstein and J.~Sonnenschein,
 JHEP {\bf 0402}, 015 (2004)
 [arXiv:hep-th/0309011].

\bibitem{Dymarsky:2009cm}
 A.~Dymarsky, S.~Kuperstein and J.~Sonnenschein,
 JHEP {\bf 0908}, 005 (2009)
 [arXiv:0904.0988 [hep-th]].

\bibitem{Bena:2011wh}
 I.~Bena, G.~Giecold, M.~Grana, N.~Halmagyi and S.~Massai,
 arXiv:1106.6165 [hep-th].

\bibitem{Klebanov:1998hh}
 I.~R.~Klebanov and E.~Witten,
 Nucl.\ Phys.\  B {\bf 536}, 199 (1998)
 [arXiv:hep-th/9807080].

\bibitem{Klebanov:2000nc}
 I.~R.~Klebanov and A.~A.~Tseytlin,
 Nucl.\ Phys.\  B {\bf 578}, 123 (2000)
 [arXiv:hep-th/0002159].

\bibitem{Klebanov:2000hb}
 I.~R.~Klebanov and M.~J.~Strassler,
 JHEP {\bf 0008}, 052 (2000)
 [arXiv:hep-th/0007191].

\bibitem{Gauntlett:2004yd}
 J.~P.~Gauntlett, D.~Martelli, J.~Sparks and D.~Waldram,
 Adv.\ Theor.\ Math.\ Phys.\  {\bf 8}, 711 (2004)
 [arXiv:hep-th/0403002].

\bibitem{Martelli:2004wu}
 D.~Martelli and J.~Sparks,
 Commun.\ Math.\ Phys.\  {\bf 262}, 51 (2006)
 [arXiv:hep-th/0411238].

\bibitem{Maldacena:2000yy}
 J.~M.~Maldacena and C.~Nunez,
 Phys.\ Rev.\ Lett.\  {\bf 86}, 588 (2001)
 [arXiv:hep-th/0008001].

\bibitem{Lust:2008zd}
 D.~Lust, F.~Marchesano, L.~Martucci and D.~Tsimpis,
 JHEP {\bf 0811}, 021 (2008)
 [arXiv:0807.4540 [hep-th]].

\bibitem{Grana:2004bg}
 M.~Grana, R.~Minasian, M.~Petrini and A.~Tomasiello,
 JHEP {\bf 0408}, 046 (2004)
 [arXiv:hep-th/0406137].

\bibitem{Giddings:2001yu}
 S.~B.~Giddings, S.~Kachru and J.~Polchinski,
 Phys.\ Rev.\  D {\bf 66}, 106006 (2002)
 [arXiv:hep-th/0105097].

\bibitem{Benini:2006hh}
 F.~Benini, F.~Canoura, S.~Cremonesi, C.~Nunez and A.~V.~Ramallo,
 JHEP {\bf 0702}, 090 (2007)
[arXiv:hep-th/0612118].

\bibitem{Herzog:2004tr}
 C.~P.~Herzog, Q.~J.~Ejaz and I.~R.~Klebanov,
 JHEP {\bf 0502}, 009 (2005)
 [arXiv:hep-th/0412193].

\bibitem{Bergshoeff:2001pv}
 E.~Bergshoeff, R.~Kallosh, T.~Ortin, D.~Roest and A.~Van Proeyen,
 Class.\ Quant.\ Grav.\  {\bf 18}, 3359 (2001)
 [arXiv:hep-th/0103233].

\bibitem{Bigazzi:2008zt}
 F.~Bigazzi, A.~L.~Cotrone and A.~Paredes,
 JHEP {\bf 0809}, 048 (2008)
 [arXiv:0807.0298 [hep-th]].

\bibitem{Nunez:2010sf}
 C.~Nunez, A.~Paredes and A.~V.~Ramallo,
 Adv.\ High Energy Phys.\  {\bf 2010}, 196714 (2010)
 [arXiv:1002.1088 [hep-th]].

\bibitem{Koerber:2007hd}
 P.~Koerber and D.~Tsimpis,
 JHEP {\bf 0708}, 082 (2007)
 [arXiv:0706.1244 [hep-th]].

\end{thebibliography}

\end{document}